# Effect of Magnetoelectrical Interactions on the Multiferroic Domain Walls

Z. V. Gareeva[a, b, *] and A. K. Zvezdin[a, c, **]

[a] *Prokhorov General Physics Institute, Russian Academy of Sciences,*
*ul. Vavilova 38, Moscow, 119991 Russia*

[b] *Institute of Molecule and Crystal Physics, Ufa Research Center, Russian Academy of Sciences,*
*pr. Oktyabrya 151, Ufa, 450075 Bashkortostan, Russia*

*\* e-mail: gzv@anrb.ru*

[c] *Ikerbasque, Basque Foundation for Science, Alameda Urquijo 36-5,*
*Plaza Bizkaia, Bilbao, Bizkaia, 48011 Spain*

*\*\* e-mail: zvezdin@gmail.com*

**Abstract**—The antiferromagnetic domain structure of a multiferroic has been investigated in the presence of a ferroelectric domain structure. It has been demonstrated that an inhomogeneous magnetoelectric (flexomagnetoelectric) interaction leads to pinning of antiferromagnetic domain walls at the walls of the ferroelectric domains and to a change in the structure of antiferromagnetic domain walls.

## 1. INTRODUCTION

Nowadays, interactions between magnetic and ferroelectric domain walls in materials with two order parameters (magnetic and electric) have been investigated extensively [1–21]. Theoretical and experimental investigations have showed that the magnetic and ferroelectric domain structures in multiferroics closely interact to one other. Theoretical studies and, subsequently, experimental investigations [22–25] have demonstrated that a nonuniform distribution of the magnetic moments induces the electrical polarization in magnetoelectric media. In [26, 27], based on a symmetry analysis, the inverse effect was predicted, namely, the appearance of spontaneous magnetization at ferroelectric domain walls of multiferroics, which are materials with a violated center of inversion and odd with respect to the time reversal. A considerable amount of experimental information on the observation and study of the ferroelectric and magnetic structures in magnetoelectric materials has been accumulated to date [1–5, 9–21]. A correlation between the magnetic and electric order parameters leads to the fact that the physical properties and characteristics of the domain structures of multiferroics differ from the properties of domain structures of antiferromagnets and ferroelectrics [1–21, 27–30]. The calculations presented in [17–19] showed that the antiferromagnetic domain structure of magnetoelectric materials, namely, hexagonal manganites, is stabilized by elastic lattice stresses at the expense of the interaction with the electric order parameter. The experimental investigations [1–3, 10–12, 15–21] also demonstrated that the ferroelectric domain structure is accompanied by the formation of the antiferromagnetic domain structure. It is customary to use the Dzyaloshinskii–Moriya magnetoelectric interaction as the main mechanism responsible for the interaction between the magnetic and electric order parameters. The specific features of the antiferromagnetic domain walls investigated in [17–19] are explained in the framework of this mechanism. At the same time, in magnetoelectric materials, other mechanism is important: the mechanism of the inhomogeneous magnetoelectric interaction [23], namely, flexomagnetoelectric interaction that is linear with respect to the magnetization gradient. This interaction has a relativistic nature, is described by the Lifshitz invariant, and plays a decisive role in investigations of incommensurate phases and phase transitions in multiferroics [31–35].

In this work, we study the antiferromagnetic domain structure of a multiferroic of the $BiFeO_3$ type, in which the Curie temperature of the electric ordering exceeds the Néel temperature. It has been demonstrated that the presence of the ferroelectric domain structure substantially affects the magnetic moment distribution; namely, the antiferromagnetic domain walls are pinned at the ferroelectric boundaries, and the direction and character of rotation of the antiferromagnetism vector are determined by the direction of the polarization vector in the ferroelectric domains. The reason of the effects revealed is an inhomogeneous magnetoelectric interaction.

## 2. FORMULATION OF THE PROBLEM AND BASIC EQUATIONS

(1) Multiferroics are materials combining two types of ordering, such as the ferroelectric and antiferromagnetic ordering. It is natural to represent a multiferroic as two interacting magnetic and electric subsystems. The free energy of a multiferroic involves the energies of the magnetic, electric, and magnetoelectric interactions

$$F = F_m + F_{el} + F_{me}, \quad (1)$$

where the magnetic interaction energy $F_m$ is represented as an expansion in the order parameter, for which, in this case, two vectors are used: the ferromagnetism vector **M** and the antiferromagnetism vector **L**; then, the energy $F_m$ takes the form

$$F_m = \frac{a_1 m^2}{2} + \frac{b_1 l^2}{2} + \frac{b_2 l^4}{2} + \frac{d}{2}(\mathbf{ml})^2$$
$$- \mathbf{MH} + F_{anis}(\mathbf{l}) + F_{exch}(\mathbf{l}).$$

Here, $a_1$, $b_1$, $b_2$, and $d$ are the expansion constants, $\mathbf{m} = \frac{\mathbf{M}}{|M|}$ and $\mathbf{l} = \frac{\mathbf{L}}{|L|}$ are the unit vectors of the ferromagnetism and antiferromagnetism, respectively, **H** is the applied magnetic field,

$$F_{exch}(\mathbf{l}) = A\left(\frac{\partial l_i}{\partial x_j}\right)\left(\frac{\partial l_i}{\partial x_j}\right)$$

is the energy of the inhomogeneous exchange interaction, $A$ is the exchange interaction constant, and $F_{anis}(\mathbf{l})$ is the magnetic anisotropy energy.

In what follows we consider epitaxial BiFeO$_3$ films, in which a growth anisotropy is induced [36, 37]. In many cases, the growth-induced anisotropy of bismuth ferrite films can be represented as the orthorhombic magnetic anisotropy [38]: $F_{anis} = K_1(\mathbf{nl})^2 + K_2(\mathbf{n}_\perp \mathbf{l})^2$, where $K_1$ and $K_2$ are the orthorhombic anisotropy constants and $\mathbf{n} = \frac{\mathbf{l}}{l}$ is the unit vector directed along **l**. Moreover, as is in the antiferromagnetism theory [39], we can, using an isomodulus approximation and condition (**ml**) = 0, exclude the vector **m** from the free energy $F_m$ remaining the dependence of the total energy of the magnetic subsystem only on the parameter **l**.

We represent the free energy of the electric subsystem, according to the theory of ferroelectricity [40], as an expansion in the order parameter η which is determined by the crystal symmetry

$$F_{el} = \frac{a_1 \eta^2}{2} + \frac{a_2 \eta^4}{4} + \dots.$$

Here, $a_1$ and $a_2$ are the expansion constants. For sake of simplicity, we choice the polarization **P** as the parameter η. A specific form of the expansion of $F_{el}(\eta)$ in the order parameter is of no importance in our case. The fact that the spontaneous polarization **P** is realized in the system at temperatures below the Curie point ($T < T_c$) is of importance for this problem. To satisfy this condition, we use the function $F_{el}(\eta)$ presented above.

The magnetoelectric interaction involves two terms linear in the electric polarization

$$F_{me} = F_{me}^1 + F_{me}^2,$$

where

$$F_{me}^1 = D_1 \mathbf{P}[\mathbf{l} \times \mathbf{m}]$$

is the Dzyaloshinskii–Moriya magnetoelectric interaction (**P** is the unit vector directed along the electric polarization), and

$$F_{me}^2 = D_2 \mathbf{P}\{(\mathbf{l}\mathrm{grad})\mathbf{l} - \mathbf{l}\mathrm{div}\mathbf{l}\} = -D_2\{\mathbf{P}(\mathbf{l}\mathrm{rot}\mathbf{l}) + \mathbf{P}\mathbf{l}\mathrm{div}\mathbf{l}\}$$

is the nonuniform magnetoelectric interaction which also called the flexomagnetoelectric interaction [35]. Here, $D_1$ is the magnetoelectric interaction constant and $D_2$ is the flexomagnetoelectric interaction constant.

With allowance of aforementioned assumptions, we can rewrite the total energy of a multiferroic (1) as

$$F = -\frac{\chi_\perp}{2}(\mathbf{H}_{eff}^2 - (\mathbf{H}_{eff}\mathbf{n})^2) + F_{exch}(\mathbf{l})$$
$$+ F_{anis}(\mathbf{l}) + F_{me}^2 + \frac{a_1 P_z^2}{2} + \frac{a_2 P_z^4}{4}, \quad (2)$$

where $\mathbf{H}_{eff} = \mathbf{H} + \mathbf{H}_{me}$ is the effective magnetic field acting to spins of the antiferromagnetic subsystem, and $\mathbf{H}_{me} = D_1[\mathbf{n} \times \mathbf{P}]$ is the magnetoelectric field acting to the antiferromagnetic subsystem and trying to induce a slightly ferromagnetic moment. The physical sense of the first term is clear: it reflects a strong anisotropy of the susceptibility of the antiferromagnetic subsystem which is determined by $\chi_\perp$ ($\chi_\perp \sim 10^{-5}$). Expression (2) is derived in [32, 39] on the assumption that the system is far from the Néel point ($T \ll N_N$)) and, because of this the longitudinal magnetic susceptibility $\chi_\parallel \ll \chi_\perp$ is neglected. In the calculations that follow, this assumption is of no principal importance.

To elucidate the main specific features of the multiferroic domain structure, we consider a mathematically simple and physically sapid special case when an easy plane (coinciding with the (111) plane of the crystal) type situation is realized. The magnetoelectric effect in bismuth ferrite films with such an orientation was investigated, for example, in [36, 37]. In the special case under consideration, the problem can be solved analytically and can be elucidated new pecu-liarities inherent in the domain structure of a multiferroic.

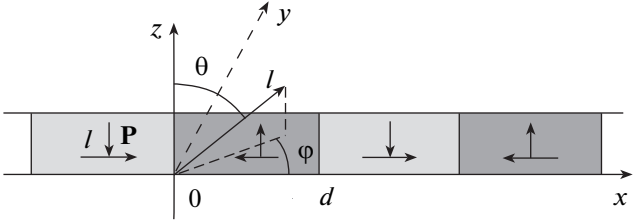

**Fig. 1.** Periodic domain structure of the multiferroic under consideration. **P** is the polarization vector, **l** is the antiferromagnetism vector, and $d$ is the domain size. The sizes of the ferroelectric and antiferromagnetic domains coincide. The directions of the polarization and antiferromagnetism vectors in the adjacent domains are opposite.

To solve the problem formulated above, we go to the angular variables $\mathbf{l} = (\sin\theta\cos\varphi, \sin\theta\sin\varphi, \cos\theta)$, the angle $\theta$ is measured from the positive direction of axis $OZ \parallel C_3 \parallel \langle 111 \rangle$, and the angle $\varphi$ lies in the film plane (Fig. 1).

In the angular variables $\theta$ and $\varphi$, the free energy density (2) is written as

$$F = A\left[\left(\frac{d\theta}{dx}\right)^2 + \sin^2\theta\left(\frac{d\varphi}{dx}\right)^2\right]$$
$$+ \left(K_1 - \frac{\chi_\perp D_1}{2}P_z^2(x)\right)\sin^2\theta + K_2\sin^2\theta\cos^2\varphi \quad (3)$$
$$- D_2 P_z(x)\left(\cos\varphi\frac{d\theta}{dx} - \sin\theta\cos\theta\sin\varphi\frac{d\varphi}{dx}\right).$$

It follows from relationship (3) that, at $K_{1\text{eff}} = K_1 - \frac{\chi_\perp D_1}{2}P_z^2 < 0$ and $K_2 < 0$, the structure in which the spins are in the $(\theta_0 = \pi/2, \varphi_0 = 0, \pi)$ plane is realized. Assume that $P(x)$ is a step function (with a jump at the ferroelectric domain wall). This approximation is used because the characteristic thickness of the magnetic domain walls of multiferroics, which is changed from 10 nm to 100 nm [1, 28–30], significantly excesses the thickness of the ferroelectric domain walls of an order of 1 nm and smaller [1, 4, 5]. We determine the domain-wall structure from the system of the Euler–Lagrange equations which for the functional (3) has the form

$$\frac{d^2\theta}{d\xi^2} - \left\{\left(\frac{d\varphi}{d\xi}\right)^2 + \kappa - \cos^2\varphi\right\}\sin\theta\cos\theta$$
$$+ 2\varepsilon_0\varepsilon(\xi)\sin\varphi\frac{d\varphi}{d\xi}\sin^2\theta = \varepsilon_0\frac{d\varepsilon}{d\xi}\cos\varphi, \quad (4)$$

$$\frac{d}{d\xi}\left\{\sin^2\theta\frac{d\varphi}{d\xi}\right\} - \sin^2\theta\sin\varphi\cos\varphi$$
$$- 2\varepsilon_0\varepsilon(\xi)\sin\varphi\frac{d\theta}{d\xi}\sin^2\theta = -\varepsilon_0\frac{d\varepsilon}{d\xi}\sin 2\theta\sin\varphi. \quad (5)$$

Here, we use the designations

$$\xi = \frac{x}{\Delta}, \quad \Delta = \sqrt{\frac{A}{|K_2|}}, \quad \varepsilon_0 = \frac{D_2|P_z|}{2\sqrt{A|K_2|}}, \quad \kappa = \frac{-|K_{1\text{eff}}|}{|K_2|},$$

and $\varepsilon(\xi)$ is a periodic function with the period $2d/\Delta$ which, in the range $[-d/\Delta, d/\Delta]$, takes values

$$\varepsilon(\xi) = \begin{cases} 1, & 0 < \xi < \frac{d}{\Delta}, \\ -1, & -\frac{d}{\Delta} < \xi < 0, \end{cases}$$

$$\frac{d\varepsilon}{d\xi} = \sum_{n=-\infty}^{\infty} (-1)^n \delta\left(\xi - \frac{d}{2\Delta}(2n+1)\right).$$

Now, we consider the process of constructing solutions to Eqs. (4) and (5) in the region occupied by one ferroelectric domain $[0, d/\Delta]$. By integration of Eqs. (4) and (5) with respect to $\xi$ over an infinitely small vicinity of the ferroelectric domain interfaces $[-\alpha, \alpha]$ and $[d/\Delta - \alpha, d/\Delta + \alpha]$, where $\alpha \longrightarrow 0$, we find the boundary conditions for Eqs. (4) and (5)

$$\left.\frac{d\theta}{d\xi}\right|_{\xi=\alpha} - \left.\frac{d\theta}{d\xi}\right|_{\xi=-\alpha} = \varepsilon_0\cos\varphi(0),$$
$$\left.\frac{d\theta}{d\xi}\right|_{\xi=d/\Delta+\alpha} - \left.\frac{d\theta}{d\xi}\right|_{\xi=d/\Delta-\alpha} = \varepsilon_0\cos\varphi\left(\frac{d}{\Delta}\right),$$
$$\left.\frac{d\varphi}{d\xi}\right|_{\xi=\alpha} - \left.\frac{d\varphi}{d\xi}\right|_{\xi=-\alpha} = \varepsilon_0\cot\theta(0)\sin\varphi(0), \quad (6)$$
$$\left.\frac{d\varphi}{d\xi}\right|_{\xi=d/\Delta+\alpha} - \left.\frac{d\varphi}{d\xi}\right|_{\xi=d/\Delta-\alpha} = \varepsilon_0\cot\theta\left(\frac{d}{\Delta}\right)\sin\varphi\left(\frac{d}{\Delta}\right).$$

(2) We represent the solution of the system of Eqs. (4)–(6) as a series of the perturbation theory with the small parameter $\varepsilon_0$:

$$\theta = \theta_0 + \varepsilon_0\theta_1 + \ldots,$$
$$\varphi = \varphi_0 + \varepsilon_0\varphi_1 + \ldots. \quad (7)$$

As aforementioned, we consider that $\theta_0 = \pi/2$; then, a zeroth approximation for the angle $\varphi$ is found by solu-

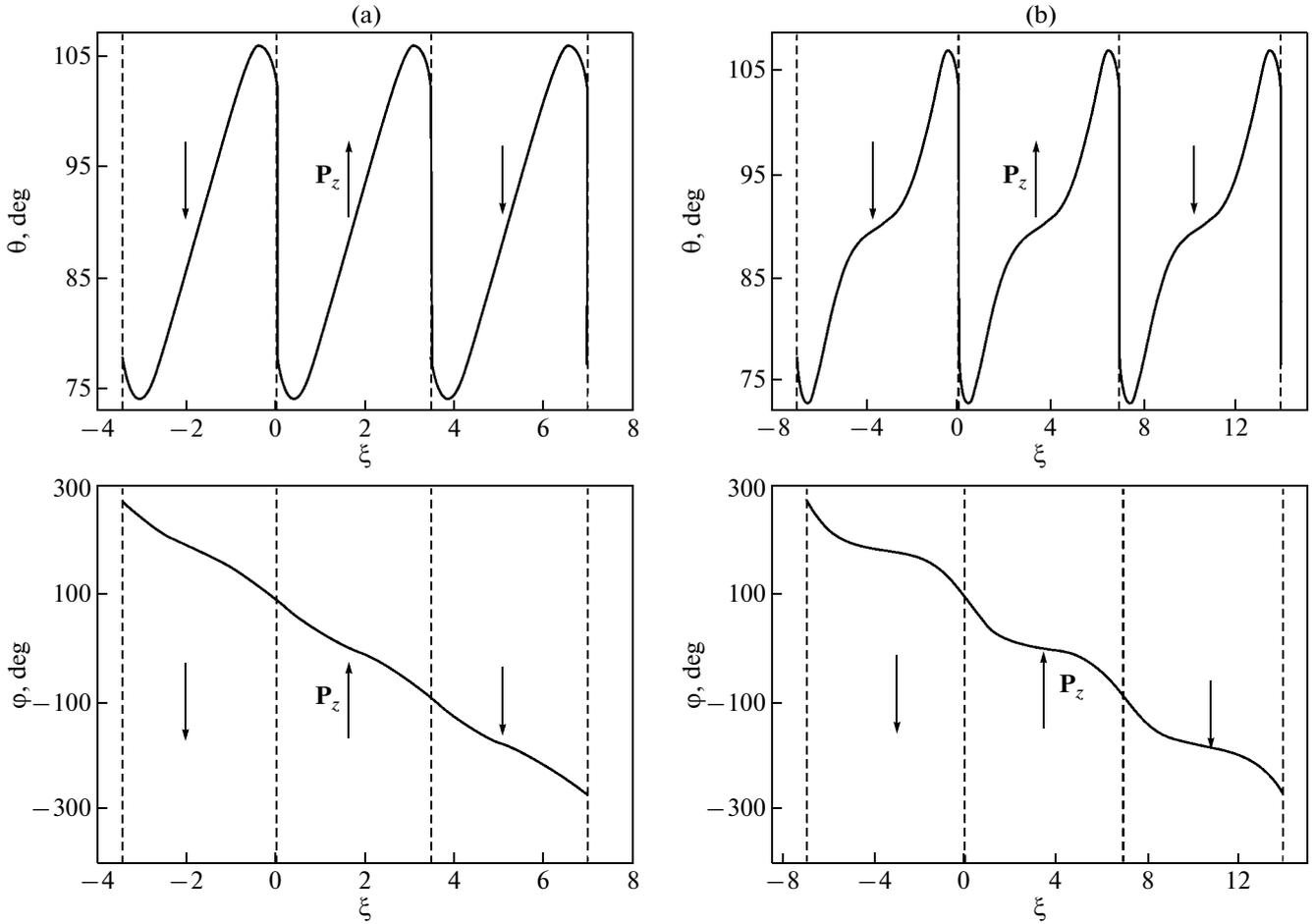

**Fig. 2.** Dependences of the angles $\theta$ and $\varphi$ on the parameter $\xi$ for (a) $d = 20$ nm ($d/\Delta = 3.46$), (b) $d = 40$ nm ($d/\Delta = 6.92$), and (c) $d = 60$ nm ($d/\Delta = 10.39$). $A = 2 \times 10^{-7}$ erg/cm, $|K_1| = 2 \times 10^6$ erg/cm$^3$, $|K_2| = 6 \times 10^5$ erg/cm$^3$, $P_0 = 6 \times 10^{-5}$ C/cm$^2$, $\lambda = 62 \times 10^{-7}$ cm, $D_2 = \dfrac{4\pi A}{\lambda P_0}$ ($\kappa = |K_1|/|K_2| = 3.3$, $\eta = \dfrac{D_2 P_0}{2\sqrt{A|K_2|}} = 0.58$; $\Delta = 5.7 \times 10^{-7}$ cm).

tion of the equation $\dfrac{d^2\varphi_0}{d\xi^2} - \sin\varphi_0 \cos\varphi_0 = 0$. The solution is

$$\varphi_0 = \frac{\pi}{2} + \arcsin\left(\operatorname{sn}\left(\pm\frac{\xi - \xi_0}{m}, m\right)\right), \quad (8)$$

where $\operatorname{sn}\left(\pm\dfrac{\xi - \xi_0}{m}, m\right)$ is the Jacobi elliptic function, and $\xi_0$ and $m$ are free parameters which can be found from additional conditions.

(3) Let us consider the case when the periods of the antiferromagnetic domain structure and ferroelectric structure coincide. Then, the modulus of the elliptic function $m$ can be found from the condition of the periodicity of the antiferromagnetism vector **l** in the plate under consideration

$$\frac{2d}{\Delta m} = K(m). \quad (9)$$

Here, $d$ is the ferroelectric domain width and $K(m) = \displaystyle\int_0^{\pi/2} \dfrac{d\varphi}{\sqrt{1 - m^2 \sin^2\varphi}}$ is the total first-order elliptical integral.

(4) The parameter $\xi_0$ determines the localization of the antiferromagnetic domain walls with respect to the ferroelectric domain structure. We assume that this parameter is such that the centers of the ferroelectric and antiferromagnetic domain structures coincide. In this case, $\xi_0$ takes the value $\xi_0 = 0$. We show further that this solution is actually determined from the condition of the domain-wall energy minimum.

(5) We consider the question as to the form of the multiferroic domain structure. Consider the zeroth

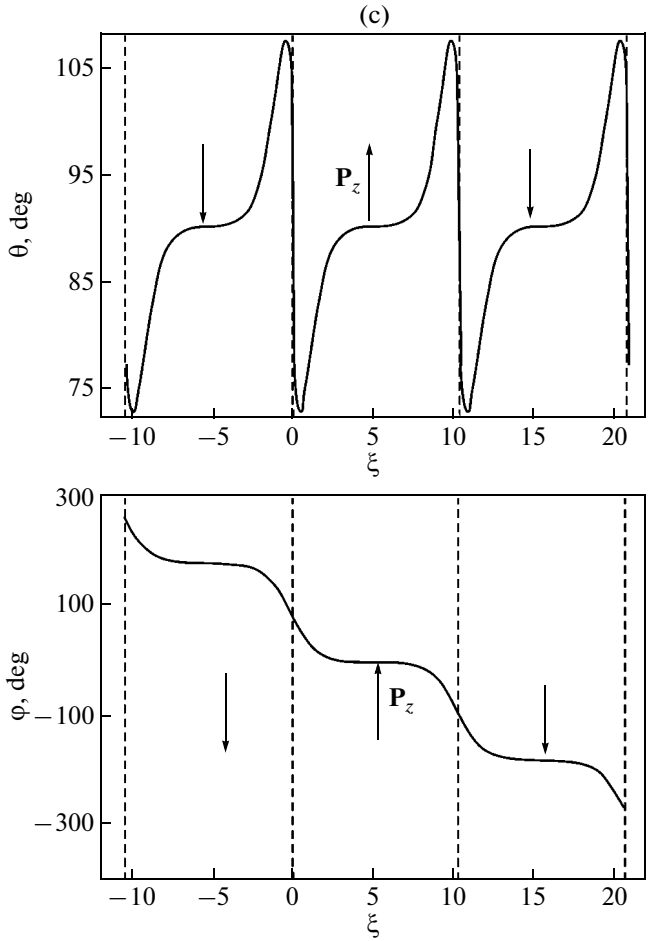

**Fig. 2.** (Contd.)

approximation with respect to the parameter $\varepsilon_0$ in the form

$$\varphi_0 = \frac{\pi}{2} - \arcsin\left(\operatorname{sn}\left(\frac{\xi - \xi_0}{m}, m\right)\right). \quad (10)$$

In what follows, we will show that exactly the chosen solution of Eq. (10) corresponds to the condition of the domain-wall energy minimum. Determine the first correction to the angle $\theta$ from the value of $\varepsilon_0$ using the equation

$$\frac{d^2\theta_1}{d\xi^2} + \left\{\left(\frac{d\varphi_0}{d\xi}\right)^2 + \kappa - \cos^2\varphi_0\right\}\theta_1 \\ + 2\sin\varphi_0\frac{d\varphi_0}{d\xi} = \cos\varphi_0\delta\left(\xi - \frac{d}{2\Delta}\right), \quad (11)$$

with the boundary conditions

$$\left.\frac{d\theta_1}{d\xi}\right|_{\xi = \alpha} - \left.\frac{d\theta_1}{d\xi}\right|_{\xi = -\alpha} = \frac{1}{2}(\cos\varphi_0)\Big|_{\xi = 0} = \frac{1}{2}\operatorname{sn}\left(\frac{\xi_0}{m}, m\right),$$

$$\left.\frac{d\theta_1}{d\xi}\right|_{\xi = d/\Delta + \alpha} - \left.\frac{d\theta_1}{d\xi}\right|_{\xi = d/\Delta - \alpha} = \frac{1}{2}(\cos\varphi_0)\Big|_{\xi = d/\Delta} \quad (12)$$

$$= -\frac{1}{2}\operatorname{sn}\left(\frac{\frac{d}{\Delta} - \xi_0}{m}, m\right).$$

The first correction to the angle $\varphi$ is found from the equation

$$\frac{d^2\varphi_1}{d\xi^2} - \varphi_1\cos 2\varphi_0 = 0, \quad (13)$$

with the boundary conditions

$$\left.\frac{d\varphi_1}{d\xi}\right|_{\xi = \alpha, d/\Delta + \alpha} - \left.\frac{d\varphi_1}{d\xi}\right|_{\xi = -\alpha, d/\Delta - \alpha} \\ = \frac{1}{2}\cot\theta_0\sin\varphi_0 = 0. \quad (14)$$

The subsequent corrections are determined similarly. We investigate the character of rotation of the antiferromagnetism vector assuming that $\xi_0 = 0$. Substituting the solution of Eq. (10) and corrections to the angles $\theta$ and $\varphi$ calculated by Eqs. (11)–(14) into relationship (7) and find the $\theta$ and $\varphi$ angles as functions of the coordinate $\xi$. The $\theta(\xi)$ and $\varphi(\xi)$ graphs calculated for different thickness of the ferroelectric domain $d$ are shown in Fig. 2. It is seen from the graphs that a certain distribution of the antiferromagnetism vector corresponds to a given ferroelectric structure. In the other words, the antiferromagnetism vector direction is related to the polarization vector direction in the ferroelectric domains. The rotation through the angle $\varphi$ begins from $\varphi(\xi = d/2\Delta) = 0$ in the domains with positive direction of the vector **P** and from $\varphi(\xi = -d/2\Delta) = \pi$ in the domains with negative **P**. The angle $\theta$ is deviated from the $\theta(\xi = -d/2\Delta, d/2\Delta) = \pi/2$, which indicates the emergence of spins from the rotation plane *XOY*. A new moment characterizing the distribution of spins in the domain walls of multiferroics is the fact that the flexomagnetoelectric interaction causes the emergence of spins from the plane of rotation when approaching the ferroelectric domain wall. It differentiates the domain walls of multiferroics from the traditional antiferromagnetic domain walls and, in essence, the flexomagnetoelectric interaction is exactly the reason of pinning.

(6) We consider the domain-wall energy and show that the energy minimum is really realized at the localization parameter equal to $\xi_0 = 0$. We find the parameter $\xi_0$ from the condition of the minimum of the domain-wall energy. The energy of the antiferromagnetic domain wall is determined as

$$\sigma_{DW} = \int_0^d \left\{ A\left[\left(\frac{d\theta}{dx}\right)^2 + \sin^2\theta\left(\frac{d\varphi}{dx}\right)^2\right] \right.$$
$$\left. -\left(|K_1| - \frac{\chi_\perp D_1}{2}P_z^2\right)\sin^2\theta - |K_2|\sin^2\theta\cos^2\varphi \right\} dx \quad (15)$$
$$- \int_0^d D_2 P_z \left(\cos\varphi\frac{d\theta}{dx} - \sin\theta\cos\theta\sin\varphi\frac{d\varphi}{dx}\right) dx.$$

The system is stable at certain value of the parameter $\xi_0$ corresponding to the antiferromagnetic domain-wall energy minimum. Find the equilibrium values of the parameter $\xi_0$. Consider the zero approximation with respect to the parameter $\varepsilon_0$ in the form of relationship (10). Find $\theta_1(\xi, \xi_0)$ and $\varphi_1(\xi, \xi_0)$ from Eqs. (11)–(14). We substitute the solution of Eq. (7) with inclusion of the correction found into Eq. (15) and determine the domain-wall energy as a function of the parameter $\xi_0$. Figure 3 depicts the antiferromagnetic domain-wall energy dependence $\sigma_{DW}(\xi_0)$. The value of $\xi_0$ is varied in the range $[-d/2\Delta, d/2\Delta]$. In this case, the point $\xi_0 = -d/2\Delta$ corresponds to the center of the ferroelectric domain with negative direction of the polarization vector, the point $\xi_0 = d/2\Delta$ corresponds to the center of the ferroelectric domain with positive direction of the polarization vector, and $\xi_0 = 0$ corresponds to the ferroelectric domain wall. It is seen from the graphs that the value of the parameter $\xi_0 = 0$ corresponds to the minimum of the $\sigma_{DW}(\xi_0)$ function. This fact testifies that only the antiferromagnetic domain walls localized at the ferroelectric domain walls are energetically stable. In other words, the antiferromagnetic domain walls are pinned to the ferroelectric domain walls.

Thus, we showed that the antiferromagnetic domain walls are localized at the ferroelectric domain walls (i.e., at $\xi_0 = 0$). In other words, the ferroelectric domain walls are the pinning centers for the antiferromagnetic domain walls.

(7) The effect of pinning of antiferromagnetic domain walls by the ferroelectric domain walls was observed experimentally [3, 13, 18, 20]. The physical mechanism of the phenomenon remains a subject of discussion. In the theoretical investigations [18, 19], the correlation between the antiferromagnetic and ferroelectric domain walls in the hexagonal manganites $YMnO_3$ is explained based on the microscopic Dzyaloshinskii–Moriya mechanism, namely, mechanism of the antisymmetric exchange interaction between the $Mn^{3+}$ ions. The estimations presented in [18, 19] show an energetic preference of the clamping effect, i.e., the effect of pinning of antiferromagnetic domain walls by the ferroelectric domain walls. However, in the framework of the approach used, of a substantial importance is the mutual configuration of the $Mn^{3+}$ spins determined by the group of the crystal magnetic symmetry. In particular, to explain the clamping-effect in the $ErMnO_3$ compound with other space group, the inclusion of additional mechanism is necessary. This fact shows that the problems of interaction between the electric and magnetic order parameters need additional investigation. Numerous facts presented in [35] demonstrate that, as the magnetic order parameter in the magnetoelectric media is inhomogeneous, the inhomogeneous magnetoelectric interac-tion is always manifested. Because of this, this interaction should be taken into account as the domain structure of a multiferroic is investigated. In this work, when studying the processes of interaction between magnetic and ferroelectric domain walls, we took into account to mechanisms of the magnetoelectric interactions linear in polarization: the homogeneous Dzyaloshinskii–Moriya mechanism and inhomogeneous flexomagnetoelectric mechanism. As the calculation show, in the problem considered, the flexomagnetoelectric interaction plays a main role in explanation of the pinning effect of the antiferromagnetic domain walls.

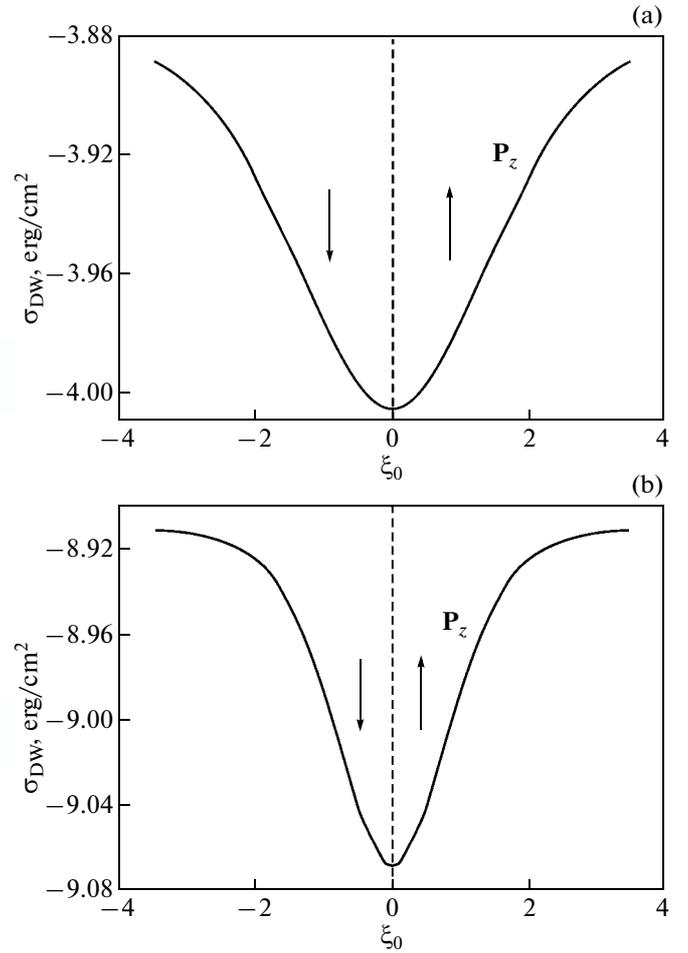

**Fig. 3.** Domain-wall energy $\sigma_{DW}$ as a function of the parameter $\xi_0$ determining the localization of the antiferromagnetic domain walls for (a) $d = 20$ nm ($d/\Delta = 3.46$) and (b) $d = 40$ nm ($d/\Delta = 6.92$). $A = 2 \times 10^{-7}$ erg/cm, $|K_1| = 2 \times 10^6$ erg/cm$^3$, $|K_2| = 6 \times 10^5$ erg/cm$^3$, $P_0 = 6 \times 10^{-5}$ C/cm$^2$, $\lambda = 62 \times 10^{-7}$ cm, $D_2 = \frac{4\pi A}{\lambda P_0}$ ($\kappa = |K_1|/|K_2| = 3.3$, $\eta = \frac{D_2 P_0}{2\sqrt{A|K_2|}} = 0.58$; $\Delta = 5.7 \times 10^{-7}$ cm).

On the whole, the study performed in this work shows that the ferroelectric domain structure cardinally changes the spin density distribution in the antiferromagnetic domain wall. The magnetoelectric interactions in multiferroics hamper the magnetic domain structure as follows: the antiferromagnetic domain walls are pinned on the ferroelectric domain walls. The spin distribution in the domain walls becomes more complex. Because of the flexomagnetoelectric interaction, the distribution of the magnetization in the plane is changed and, moreover, spins emerge from the plane near the ferroelectric domain walls. The flexomagnetoelectric interaction changes the domain-wall energy of multiferroics. In common antiferromagnets, the rotation of the antiferromagnetism vector occurs in the sample plane, and the energy of the antiferromagnetic domain wall is determined by the known relationship $\sigma_0 = 4\sqrt{AK_2}$. As shown above, in multiferroics, spins emerge from the plane during their rotation: the position of the antiferromagnetism vector is determined by two angles $\theta$ and $\varphi$, and the energy of the antiferromagnetic domain-wall energy calculated by relationship (15) differs from $\sigma_0$.

## ACKNOWLEDGMENTS

We are grateful to R.A. Doroshenko and A.P. Pyatakov for their participation in discussions of the results obtained in this work.

This study was supported by the Russian Foundation for Basic Research (project no. 08-02-01068-a).